\documentclass[aps,nofootinbib,longbibliography,prd,twocolumn]{revtex4-1}
\usepackage[babel]{csquotes}
\usepackage{graphicx}
\usepackage{amsmath,amssymb}
\usepackage[colorlinks,citecolor=blue,linkcolor=blue,urlcolor=blue]{hyperref}
\usepackage{mathrsfs}
\usepackage{enumerate}
\def\be{\begin{equation}}
\def\ee{\end{equation}}

\begin{document}

\title{The dangers of non-empirical confirmation}

\author{Carlo Rovelli}

\affiliation{\small
\mbox{CPT, Aix-Marseille Universit\'e, Universit\'e de Toulon, CNRS, F-13288 Marseille, France.} }
\date{\small\today}

\begin{abstract}

\noindent 
In the book ``String Theory and the Scientific Method" \cite{Dawid2013}, Richard Dawid describes a few of the many non-empirical arguments that motivate theoretical physicists' confidence in a theory, taking string theory as case study.  I argue that excessive reliance on non-empirical evidence compromises the reliability of science, and that precisely the case of string theory well illustrates this danger.

\vspace{1mm}

\noindent  {\em Contribution to the meeting ``Why Trust a Theory? Reconsidering Scientific Methodology in Light of Modern Physics," Munich, Dec. 7-9, 2015.}

\end{abstract}

\maketitle

\noindent Scientists have always relied on non-empirical arguments to trust theories. They choose, develop and trust theories \emph{before} finding empirical evidence.  The entire history of science witnesses for this. Kepler trusted Copernicus'  theory \emph{before} its predictions surpassed Ptolemy's; Einstein trusted General Relativity \emph{before} the detection of the bending of light from the sun.   They had non-empirical arguments, which proved good.   

According to a popular version of the Popper-Kuhn account of the scientific activity, theories are generated at random, sort of fished out from the blue sky, and then judged only on empirical ground. This account is unrealistic: theorists do not develop theories at random.  They use powerful theoretical, non-empirical, motivations for creating, choosing and developing theories.  If these did not exist, the formidable historical success of theoretical physics would be incomprehensible.  To evaluate theories, they routinely employ a vaste array of non-empirical arguments, increasing or decreasing their confidence in this or that theoretical idea, \emph{before} the hard test of empirical confirmation (on this, see Chapter VIII of \cite{Rovelli2011g}). This is the context of a ``preliminary appraisal" of theories, or ``weak" evaluation procedures \cite{Schaffner1993}.   

In the book ``String Theory and the Scientific Method" \cite{Dawid2013}, Richard Dawid describes some of these non-empirical arguments that motivate theoretical physicists' confidence in a theory, taking string theory as case study. This may imply that the use of non-empirical arguments is somewhat of a novelty in scientific practice. It is not.  

But the theorists's ``preliminary appraisal" of theories is quite another matter than the hard empirical testing of a theory, and fogging the distinction is a mistake. 

Dawid uses a Bayesian paradigm to describe how scientists evaluate theories.  Bayesian confirmation theory employs the verb ``confirm" in a technical sense which is vastly different from its common usage by lay people and scientists. In Bayesian theory, ``confirmation" indicates any evidence in favour of a thesis, \emph{however weak}.  

 In Bayesian parlance, for instance, seeing a Chinese in Piccadilly Circus ``confirms" the theory that the majority of Londoners are Chinese.  Nobody says so outside Bayesian theory.   For lay people and scientists alike, ``confirmation" means something else: it means ``very strong evidence, sufficient to accept a belief as reliable".   

This unfortunate ambiguity has played a role in the reaction to Dawid's work: some scientists appreciated his recognition of their theoretical reasons for defending a theory; but some string theorists went further: they were all too happy that string theory, which  lacks ``confirmation" (in the standard sense), was promoted by Dawid to have plenty of ``confirmation" (in Bayesian sense), raising sharp contrary reactions \cite{Ellis2014}.  

Unfortunately Dawid himself has done little to dispel this ambiguity, and this generates a problem for his views, for the following reason.  

Bayesian confirmation theory allows us to talk about the spectrum of intermediate degrees of credence between theories that are ``confirmed", in the common sense of the word, or ``established", and theories which are still ``speculative", or ``tentative".   But doing so it obfuscates precisely the divide that \emph{does} exist in science between a confirmed theory and a tentative one.  We trust the existence of the Higgs particle, which is today the weakest of the confirmed theories, with a 5-sigma reliability, namely a Bayesian degree of confidence of 99.9999\%.   In their domains of validity, classical electrodynamics or Newtonian mechanics are even far more reliable: we routinely entrust our life to them. No sensible person would entrust her life to a prediction of string theory.

The distinction is there and is clear. A philosophy of science blind to this distinction is a bad philosophy of science. It  is so important that phrasing it in terms of higher or lower Bayesian degree of belief obfuscates the point: in science we \emph{do} have theories that are ``confirmed" or ``established", which means that are extremely reliable in their domain. Then we have other theories which perhaps enjoy the confidence of some scientists, but are tentative: we wouldn't entrust to them even our life \emph{savings}.  

The distinction between reliable theories and speculative theories may not always be perfectly sharp, but is an essential ingredient of science.  As Thoreau puts it: ``To know that we know what we know, and to know that we do not know what we do not know, that is true knowledge" \cite{Thoreau1854}.   The very existence of reliable theories is what makes science valuable to society.  Loosing this from sight is not understanding why science matters. 

Why is this relevant for non-empirical confirmation? Because non-empirical evidence is emphatically insufficient to increase the confidence of a theory to the point where we can consider it established; that is, to move it from ``maybe" to ``reliable". 

The reason only \emph{empirical} evidence can grant ``confirmation" in the common sense of the word, is crucial and important: we all tend to be blinded by our beliefs.  We pile up non-empirical arguments in support these.  The historical success of science is grounded in the readiness to give up beloved beliefs when empirical evidence is against them.   We create theories with our intelligence, use non-empirical arguments to grow confidence in them,  but then \emph{ask nature} if they are right or wrong.  They are often wrong.  Witness --if more was need-- the recent surprise of many theorists in not finding the low-energy super-symmetric particles they expected.  

As T.\,H.\,Huxley put it: ``the great tragedy of Science is the slaying of a beautiful hypothesis by an ugly fact" \cite{Huxley}.  Tragedy, yes, but an incredibly healthy one, because this is the very source of science reliability: checking non-empirical arguments against the proof of reality.  

Perhaps no science illustrates this better than medicine. The immense success of Western medicine is largely (one is tempted to say ``almost solely") based on a single idea:  checking statistically the efficacy of the remedies used.  With this simple idea, the life expectancy of us all has more than doubled in a few centuries.  That is: by simply \emph{not} trusting non-empirical arguments. 

\vspace{2mm}

Dawid's merit is to have emphasised and analysed some of the non-empirical argument that scientists use in the ``preliminary appraisal" of theories.  His weakness is to have obfuscated the crucial distinction between this and validation: the process where a theory becomes reliable, gets accepted by the entire scientific community, and potentially useful to society. The problem with Dawid is that he fails to say that,  for this, only \emph{empirical} evidence is convincing. 

String theory, Dawid's case study, illustrates well the risk of over-relying on non-empirical confirmation and the need of empirical validation.  As Dawid notices, non-empirical argument support the credence in strings. These argument are valuable, but too weak to grant reliability. I mention one:  string theorists commonly claim that string theory has no alternatives (``the only game in town").  This is the first of Dawid's non-empirical arguments.  But as any scientist knows very well, any ``no alternative" argument holds only under a  number of assumptions, and these might turn out to be false.  In fact, not only alternatives to string theory \emph{do} exists in the real world, but these alternatives are themselves considered credible by their supporters precisely because they themselves have ``no alternative" under a \emph{different} set of assumptions!  As a theory of quantum gravity, an alternative to string theory is loop quantum gravity, considered the ``only game in town" by those who embrace it, under \emph{their} set of assumptions.  Any theory, physically correct or incorrect, has ``no alternatives" under suitable assumptions; the problem is that these assumptions may be wrong. Here we see clearly the weakness of non-empirical arguments. In science we learn something solid when something challenges our assumptions, not when we hold on to them at any cost.  

String theory is a proof of the dangers of relying excessively on non-empirical arguments. It raised great expectations thirty years ago, when it promised to compute all the parameters of the Standard Model from first principles, to derive from first principles its symmetry group $SU(3)\times SU(2)\times U(1)$ and the existence of its three families of elementary particles, to predict the sign and the value of the cosmological constant, to predict novel observable physics, to understand the ultimate fate of black holes and to offer a unique well-founded unified theory of everything. Nothing of this has come true.  String theorists, instead, have predicted a negative cosmological constant, deviations from NewtonÕs $1/r^2$ law at sub millimeters scale, black holes at CERN, low-energy super-symmetric particles, and more. All this was false. 

From a Popperian point of view, these failures do not falsify the theory, because the theory is so flexible that it can be adjusted to escape failed predictions. But from a Bayesian point of view, each of these failures \emph{decreases} the credibility in the theory, because a positive result would have increased it.   The recent failure of the prediction of supersymmetric particles at LHC is the most fragrant example. By Bayesian standards, it  lowers the degree of belief in string theory dramatically.  This is an \emph{empirical} argument. Still, Joe Polchinski, prominent string theorist, writes in \cite{Polchinski2016} that he evaluates the probability of string to be correct at 98.5\% (!). 

Scientists that devoted their life to a theory have difficulty to let it go, hanging on non-empirical arguments to save their beliefs, in the face of empirical results that Bayes confirmation theory counts as negative.  This is human.  A philosophy that takes this as an exemplar scientific attitude is a bad philosophy of science.

\providecommand{\href}[2]{#2}\begingroup\raggedright\endgroup


\begin{thebibliography}{1}

\bibitem{Dawid2013}
R.~Dawid, {\em {String Theory and the Scientific Method}}.
\newblock Cambridge University Press, Cambridge, 2013.

\bibitem{Rovelli2011g}
C.~Rovelli, {\em {The First Scientist: Anaximander and His Legacy}}.
\newblock Westholme, Chicago, 2011.

\bibitem{Schaffner1993}
K.~Schaffner, {\em {Discovery and Explanation in Biology and Medicine}}.
\newblock University of Chicago Press., Chicago, 1993.

\bibitem{Ellis2014}
G.~Ellis and J.~Silk, ``{Scientific method: Defend the integrity of physics},''
  {\em Nature} {\bf 516} (2014)  321323.

\bibitem{Thoreau1854}
H.~D. Thoreau, {\em {Walden}}.
\newblock Penguin Classics, London, 1854.

\bibitem{Huxley}
T.~H. Huxley, ``{Biogenesis and abiogenesis (1870)},'' in {\em Collected
  Essays}, ch.~8, p.~229.
\newblock Harper, New York, 1959.

\bibitem{Polchinski2016}
J.~Polchinski, ``{Why trust a theory?},''
  \href{http://arxiv.org/abs/1601.06145}{{\tt arXiv:1601.06145}}.

\end{thebibliography}

\end{document}